\documentstyle[preprint,tighten,aps,floats,psfig,epsfig,amssymb]{revtex}

\begin{document}
\draft 

\title{ 
Transitions and crossover phenomena in
fully frustrated XY systems }
\author{Martin Hasenbusch,$^{1}$ Andrea Pelissetto,$^2$ and Ettore
Vicari$\,^1$ } 
\address{$^1$ Dip. Fisica dell'Universit\`a di Pisa and
INFN, Largo Pontecorvo 2, I-56127 Pisa, Italy} 
\address{$^2$
Dip. Fisica dell'Universit\`a di Roma ``La Sapienza" and INFN, P.le
Moro 2, I-00185 Roma, Italy}

\date{\today}

\maketitle

\begin{abstract}
We study the two-dimensional fully frustrated XY (FFXY) model and two
related models, a discretization of the Landau-Ginzburg-Wilson
Hamiltonian for the critical modes of the FFXY model and a coupled
Ising-XY model, by means of Monte Carlo simulations on square lattices
$L^2$, $L\lesssim 10^3$.  We show that their phase diagram is
characterized by two very close chiral and spin transitions, at
$T_{\rm ch}>T_{\rm sp}$ respectively, of the Ising and
Kosterlitz-Thouless type. At $T_{\rm ch}$ the Ising regime sets in
only after a preasymptotic regime, which appears universal to some
extent.  The approach is nonmonotonic for most observables, with a
wide region controlled by an effective exponent $\nu_{\rm eff} \approx
0.8$.
\end{abstract}



The nature of the phase transitions in frustrated systems is of great
interest in statistical physics.  The two-dimensional 
fully frustrated XY (FFXY) model
\cite{Villain-77} is defined by the Hamiltonian
\begin{equation}
{\cal H}_{\rm FFXY} = 
- J \sum_{\langle xy\rangle} {\rm cos} (\theta_x - \theta_y + A_{xy}),
\label{ffxy}
\end{equation}
where the sum is over all nearest-neighbor pairs of a square or triangular
lattice, and $A_{xy}$ satisfy the constraint $\sum A_{xy}=\pi$ around
any plaquette.  It is experimentally 
relevant for Josephson-junction arrays in a magnetic field
\cite{ML-00}.  The angle variable $\theta_x$ corresponds to the phase
of the superconducting order parameter on each superconducting grain,
and $A_{xy}$ is the vector potential of a perpendicular magnetic field
corresponding to half a flux quantum per plaquette.

The ground state of the FFXY model presents an enlarged
O(2)$\otimes{\mathbb Z}_2$ degeneracy. The additional ${\mathbb Z}_2$
degeneracy is related to the breaking of the chiral symmetry
\cite{Villain-77}.  For each plaquette $\Pi$ we define the chirality
\begin{equation}
C_n \equiv \sum_{\langle xy \rangle \in \Pi} 
{\rm sin} (\theta_x - \theta_y + A_{xy}),
\label{chirality}
\end{equation}
where $n$ is the dual-lattice site at the center of $\Pi$.  On the
square lattice the staggered magnetization $M_C \equiv
\sum_n (-1)^{n_1 + n_2} C_n/V$ defines an order parameter, which
competes with the spin modes to determine the phase diagram of the
FFXY model.  The critical behavior has been much investigated during
the last few decades (see, e.g.,
\cite{TJ-83,GKLN-91,LGK-91,RJ-92,Lee-94,LL-94,KNKB-94,%
Olsson-95,BG-97,LL-98,CVCT-98,BD-98,LS-00,FCKF-00,%
Korshunov-02,COPS-03,OI-03,GD-05,OT-05}) using Monte Carlo (MC)
simulations, real-space renormalization-group (RG) techniques,
field-theoretical methods, etc. Moreover, several related models have
also been considered: for instance the fractional-charge
Coulomb gas, coupled Ising-XY models, coupled XY models.

In spite of all this work, the phase diagram and critical behavior
of the FFXY model are still rather controversial.
Most recent MC simulations favor the existence of two very close
transitions
\cite{Lee-94,LL-94,Olsson-95,LL-98,CVCT-98,LS-00,FCKF-00,OI-03}.  The
most likely interpretation is that the higher-temperature transition
is characterized by the onset of chiral long-range order, while spins
remain disordered. The lower-temperature transition is associated with
the breaking of the continuous symmetry and is followed by a
low-temperature phase in which spin quasi-long-range order coexists
with chiral long-range order.  The chiral transition is expected to be
in the Ising universality class, due to the scalar nature of the chiral
order parameter.  The second one should be a Kosterlitz-Thouless (KT)
transition.  This scenario is also supported by arguments based on a
kink-antikink unbinding picture \cite{Korshunov-02,OT-05}. The Ising
nature of the chiral transition has not been satisfactorily supported
by numerical simulations so far.  Most MC simulations
\cite{GKLN-91,LGK-91,RJ-92,Lee-94,LL-94,KNKB-94,BG-97,LL-98,%
BD-98,LS-00,OI-03,GD-05} have found that the behavior of the chiral
modes---both in the FFXY and in related models---is not consistent
with an Ising transition. For example, most finite-size scaling (FSS)
analyses have obtained $\nu\approx 0.8$, instead of the Ising value
$\nu=1$. There are several possible explanations.  One possibility is
that, even if the chiral order parameter is a scalar, the chiral
transition belongs to a universality class that is not the Ising
one. After all, the estimate $\nu\approx 0.8$ appears to be somewhat
universal, the same value being obtained in several different models.
A second one is that the Ising regime sets in only on large
lattices \cite{Olsson-95}: the observed behavior is only an
intermediate crossover. A third possibility is that the apparent two
transitions are a finite-size effect. On the contrary, spin and
chirality order at the same temperature.  In this case the critical
spin and chiral modes couple at criticality and give rise
to a qualitatively new critical behavior, in which  chiral modes do
not behave as spins in the Ising model.

\begin{figure}[tb]
\centerline{\psfig{width=8truecm,angle=0,file=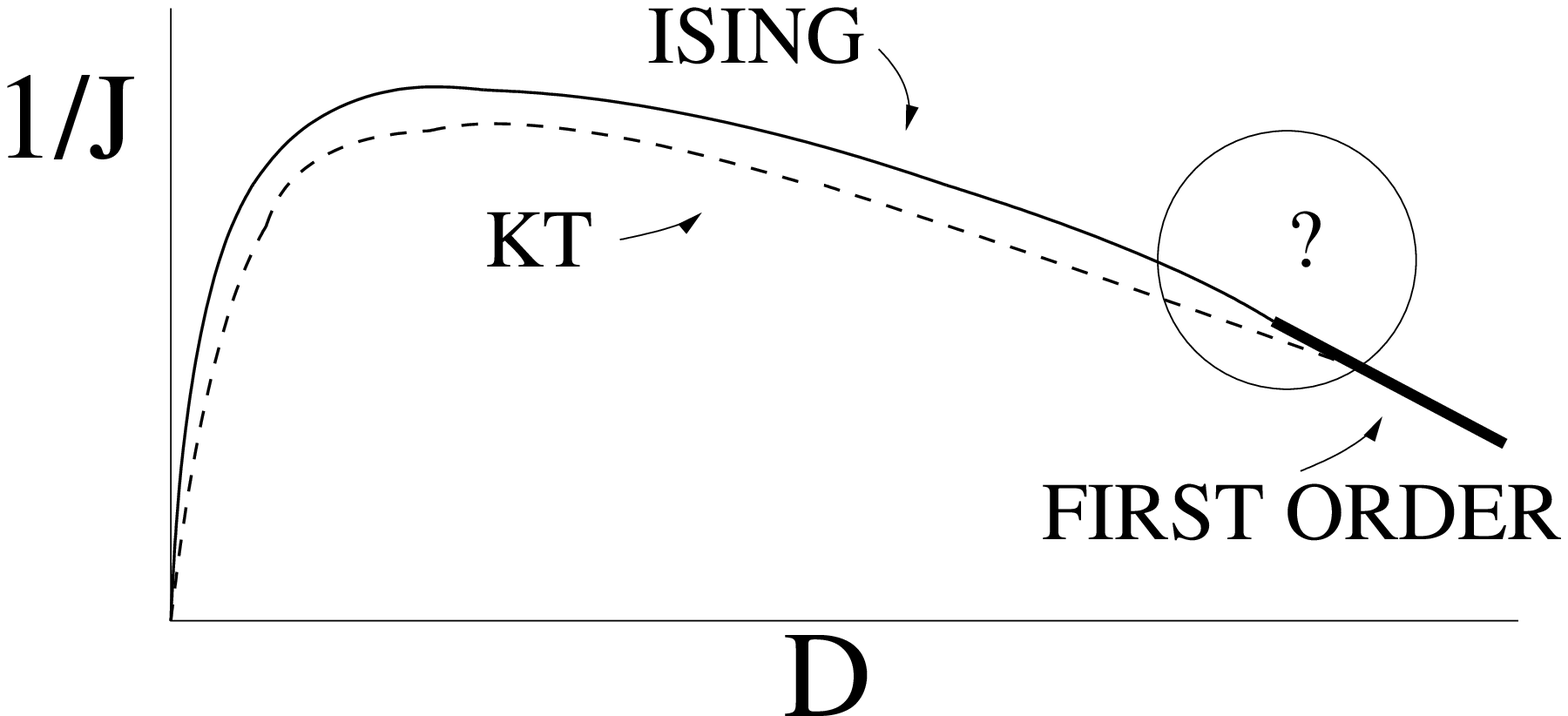}}
\vspace{1mm}
\centerline{\psfig{width=8truecm,angle=0,file=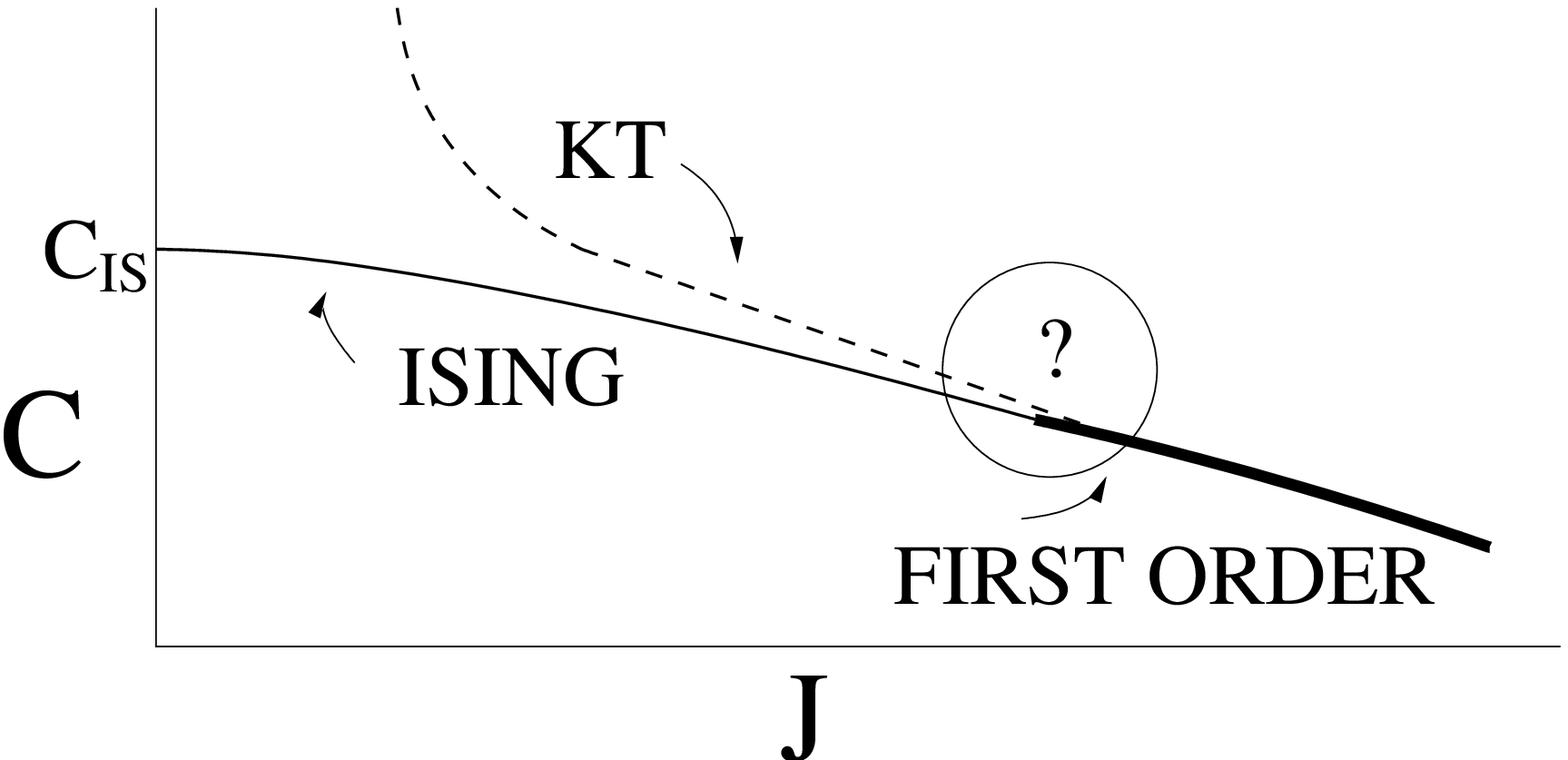}}
\vspace{1mm}
\caption{ 
Sketches of the phase diagrams of the $\phi^4$ model for $U=1$ (above)
and  of the IsXY model (below). The behavior in the 
circled region where the transition lines meet is unclear.
}
\label{phased}
\end{figure}

In this paper we consider the square-lattice FFXY model (implemented
by alternating vertical lines with ferromagnetic and antiferromagnetic
couplings) and two related lattice models. The first one is a $\phi^4$
model defined by the Hamiltonian
\begin{eqnarray}
{\cal H}_{\phi} &=& - 
J \sum_{\langle xy\rangle,i} {\phi}_{i,x}\cdot {\phi}_{i,y} +
\sum_{i,x} \left[ \phi_{i,x}^2 + U (\phi_{i,x}^2-1)^2 \right]  
\nonumber\\
&&+ 2 \, (U+D) \sum_x \phi_{1,x}^2 \phi_{2,x}^2,
\label{HLi} 
\end{eqnarray}
where ${\phi}_{1,x}, {\phi}_{2,x}$ are real
two-component variables defined on the sites $x$ of a square lattice,
$\phi^2_i\equiv {\phi}_i \cdot {\phi}_i$, and
$J,U,D>0$.  Hamiltonian ${\cal H}_{\phi}$ is
expected to describe the critical modes of the FFXY model.  It
corresponds to a straightforward lattice discretization
of the Landau-Ginzburg-Wilson theory obtained by applying
a Hubbard-Stratonovich transformation to the FFXY model, 
see, e.g., Ref.~\cite{LGK-91}.  
Here we set $U=1$; all results reported below
refer to this case.  
The symmetry $\phi_{1} \leftrightarrow
\phi_{2}$ is the analogue of the ${\mathbb Z}_2$ chiral symmetry of
the FFXY model.  The chiral order parameter is 
\begin{equation}
C_x = \phi_{1,x}^2 - \phi_{2,x}^2.
\label{chop}
\end{equation}
When $D=0$ model (\ref{HLi}) becomes O(4)
symmetric; it does not have any transition at finite temperature, but
its correlation length diverges exponentially for $J\rightarrow
\infty$, see, e.g., Ref.~\cite{PV-review}.  
We also consider the coupled Ising-XY (IsXY) model \cite{GKLN-91}
\begin{equation}
{\cal H}_{\rm IsXY} = 
- \sum_{\langle xy\rangle} 
\left[ \frac{J}{2} (1+\sigma_x \sigma_y) \,s_x \cdot s_y 
+ C \sigma_x \sigma_y \right] ,
\label{IsXY}
\end{equation}
where $\sigma_x=\pm 1$, and the two-component spins $s_x$ satisfy
$s_x\cdot s_x=1$.  Here $s_x$ and $\sigma_x$
correspond to spin and chiral variables, respectively. 
Note that, by taking the limit $U\to
\infty$ and then $D\to \infty$ in model (\ref{HLi}),
one recovers the IsXY model for $C=0$.

We performed MC simulations on $L\times L$ lattices with periodic
boundary conditions, and sizes up to $L=O(10^3)$.  We used mixtures of
Metropolis and overrelaxation updating algorithms, as proposed in
Ref.~\cite{GP-98}.  In this paper we present the main results;
details will be reported elsewhere.  We find two very close Ising and
KT transitions in the FFXY model, and in the $\phi^4$ and IsXY models
for extended ranges of the parameters $D>0$ and $C$.  No evidences of
unique continuous transitions are found.  For sufficiently large $D$
and $-C$, i.e.  $D\gtrsim 50$ and $C\lesssim -5$, we find
instead a single first-order transition.  The phase diagrams of the
$\phi^4$ and IsXY models are shown in Fig.~\ref{phased} \cite{afisxy}.
Little is known about the region where the two continuous transition
lines turn into a single first-order one.  Another interesting feature
emerges from our numerical results: the asymptotic critical behavior
at the chiral transition is observed only after a peculiar
nonmonotonic crossover regime, which appears universal to some extent.

\begin{table}[tbp]
\squeezetable
\caption{
Comparison of estimates of $\eta$ obtained from $\chi_s$, and from
$R_s$ and $\Upsilon$ using the relations valid in the XY model, such
as Eq.~(\ref{xyupsilon}), in the low-temperature phase. 
}
\label{lowt}
\begin{tabular}{lllll}
\multicolumn{1}{c}{model}&
\multicolumn{1}{c}{$J$}&
\multicolumn{1}{c}{$\eta$}&
\multicolumn{1}{c}{$\eta$ from $R_s$}&
\multicolumn{1}{c}{$\eta$ from $\Upsilon$}\\
\tableline \hline
FFXY & 2.4 & 0.1480(5)  & 0.1479(4)   & 0.14779(11) \\
     & 2.3 & 0.1750(5)  & 0.1752(4)   & 0.1758(3) \\
     & 2.26 & 0.2023(11)  & 0.2015(7)   & 0.2021(5) \\
$\phi$, $D=1/2$ & 1.50 & 0.1341(6)  & 0.1346(4)  & 0.13425(12) \\
                     & 1.48 & 0.1675(16) & 0.1672(10) & 0.1670(4) \\
$\phi$, $D=99$ & 1.6005 & 0.120(3)   & 0.1214(6) & 0.1202(5) \\
IsXY, $C=0$ & 1.52 & 0.1817(7)  & 0.1821(4)  & 0.18190(13) \\
\end{tabular}
\end{table}

In the FFXY model the square lattice can be divided into four
sublattices, so that the four sites of every plaquette belong to
different sublattices. The ground state is translation invariant within
these sublattices. We define the spin correlation function $G_s(x)$ as
the correlation between two spins in the same $L/2\times L/2$
sublattice.  Then we consider the staggered chirality correlation
function $G_c(x)$.  In the $\phi^4$ model the spin and chiral
correlation functions are defined by $G_s(x) \equiv \langle
\sum_i \phi_{i,0} \cdot \phi_{i,x} \rangle$ and $G_c(x) = \langle C_0
\, C_x \rangle_c$.  In the IsXY model we have $G_s(x) \equiv \langle
s_{0} \cdot s_{x} \rangle$ and $G_c(x) = \langle \sigma_0 \, \sigma_x
\rangle_c$.  From $G_s$ and $G_c$ we define the
susceptibilities $\chi_s$ and $\chi_c$, and second-moment correlation
lengths $\xi_s$ and $\xi_c$.  We also consider $R_{s}\equiv
\xi_{s}/L$, $R_{c}\equiv \xi_{c}/L$, the spin and chiral Binder
parameters $B_{s}$ and $B_c$, and the helicity modulus $\Upsilon$.

\begin{figure}[tb]
\centerline{\psfig{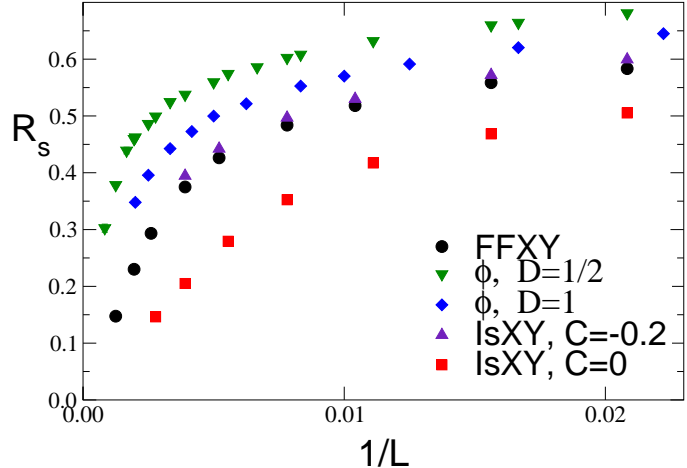}}
\caption{ 
$R_s\equiv \xi_s/L$ at fixed $R_c\equiv \xi_c/L$ vs $1/L$.
}
\label{xiols}
\end{figure}

We first show that the low-temperature phase of these models is
characterized by the breaking of the ${\mathbb Z}_2$ chiral symmetry
and by a spin quasi-long-range order analogous to the one of the
standard XY model. Indeed, chiral modes are magnetized and, in the
large-$L$ limit, the exponent $\eta$ (computed using $\chi_s\sim
L^{2-\eta}$), $R_s\equiv \xi_s/L$, and $\Upsilon$ satisfy the
universal relations that hold among the corresponding quantities in
the low-temperature phase of the XY model on a $L\times L$ square
lattice with periodic boundary conditions.  For
example~\cite{Has-05,sccorr}
\begin{equation}
\Upsilon(\eta)={1\over 2\pi\eta} - 
{2\sum_{n=-\infty}^\infty n^2 \exp (-\pi n^2/\eta)\over 
\eta^2 \sum_{n=-\infty}^\infty \exp (-\pi n^2/\eta)}
\label{xyupsilon}
\end{equation}
for $0< \eta \le 1/4$.  As shown by the results of Table~\ref{lowt},
the agreement is very good and provides a conclusive evidence that the
low-temperature phase of the FFXY and related models is controlled by
the same line of Gaussian fixed points that is relevant for the XY
model.

To determine the number of transitions, we perform a 
FSS analysis using the method proposed in
Ref.~\cite{Hasenbusch-99}, see also \cite{CHPRV-01}.  Instead of
computing the various quantities at fixed Hamiltonian parameters, we
compute them at a fixed value of a chiral---this guarantees that we
are observing the chiral transition---RG invariant quantity; in our
specific case we choose $R_c\equiv \xi_c/L$.  This method has the
advantage of not requiring a precise estimate of the critical
temperature.  We fix $R_c=R_{\rm Is}$ where $R_{\rm Is}=0.905048...$
is the corresponding Ising value \cite{SS-00}. Note that such a choice
does not represent a bias in favor of an Ising transition.  $R_c$ can
be fixed to any value $R_{c,\rm fix}$: the conclusions are independent
of $R_{c,\rm fix}$.  If the transition is unique, also the spin
correlation length $\xi_s$ diverges, and thus $R_s\equiv \xi_s/L$ at
fixed $R_c$ is expected to converge to a nonzero value.  If there are
two transitions and $T_{\rm ch}>T_{\rm sp}$, 
$\xi_s$ remains finite, so that $R_s\sim 1/L$ for
$L\rightarrow \infty$.  Results for $R_s$ are shown in
Fig.~\ref{xiols}. They appear to decrease with increasing $L$, without
showing any hint at a convergence to a nonzero value.  This shows that
the spin correlation length $\xi_s^{(c)}$ at the chiral transition is
finite, though quite large.  For example, for $L\to \infty$ we have
$\xi_s^{(c)}=118(1)$ for the FFXY model, $\xi_s^{(c)}\gtrsim 400$ for
the $\phi^4$ model with $D=1/2$, $\xi_s^{(c)}=52.7(4)$ for the IsXY
model at $C=0$. Additional evidence is provided by the helicity
modulus. At the chiral transition, it appears to vanish in the
large-$L$ limit, consistently with the fact that spin variables are
disordered.

\begin{figure}[tb]
\centerline{\psfig{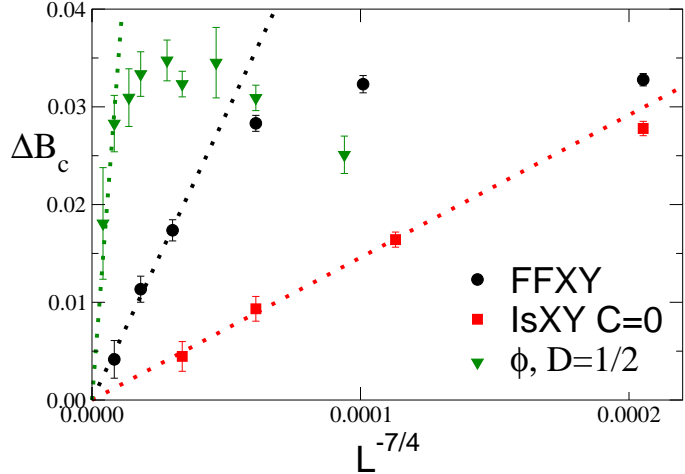}}
\vspace{1mm}
\caption{
$\Delta B_c\equiv B_c - B_{\rm Is}$ at fixed $R_c\equiv \xi_c/L$ vs $L^{-7/4}$.
}
\label{Bceffxy}
\end{figure}

As a consequence of the large value of $\xi_s^{(c)}$, the eventual
Ising regime at the chiral transition may set in only when $L\gg
\xi_s^{(c)}$ \cite{Olsson-95}.  
This fact is substantially confirmed
by our FSS analysis.  For instance, if, for any invariant quantity
$S$, we define an effective exponent $\nu_{\rm eff}(L)$ as
\begin{equation}
1/\nu_{\rm eff}(L) = \left(\ln d S/dJ|_{2L} - \ln d S/dJ|_{L}\right)/\ln 2 ,
\label{nueffdef}
\end{equation}
we find that $\nu_{\rm eff}(L)$ shows first a plateau at $\nu_{\rm
eff}(L)\approx 0.8$---this is consistent with previous works on small
lattices---and only for $L\gtrsim \xi_s^{(c)}$ appears to approach the
Ising value. Similar crossover effects are observed in other
quantities.  For example, the chiral Binder parameter $B_c$ is
systematically higher than the Ising value \cite{SS-00} $B_{\rm
Is}=1.167923(5)$.  However, as $L$ increases, $B_c$ starts to decrease
and, if $L$ is large enough, its behavior is compatible with the
expected one \cite{CCCPV-00}, $B_c = B_{\rm Is} + a L^{\eta_{\rm Is} -
2} = B_{\rm Is} + a L^{-7/4}$, see, e.g.,  Fig.~\ref{Bceffxy}.
This confirms the Ising nature of the chiral transition.

The spin transition, whenever continuous, is expected to be a KT one.
We verify that at the KT transition  
$R_s$ and $\Upsilon$ behave as expected, for example \cite{Has-05}
\begin{equation}
\Upsilon = 0.63650817819... + \frac{0.318899454...}
{\ln L+ c } + ...
\label{Yxy} 
\end{equation}
In most cases, including the FFXY model, the chiral and the spin
transitions are very close.  If $\delta\equiv (J_{\rm sp}-J_{\rm
ch})/J_{\rm ch}$, $J_{\rm sp}$ and $J_{\rm ch}$ being the location of
the two transitions, we find $\delta=0.0159(2)$ for the FFXY model
[$J_{\rm ch} = 2.2063(1)$ and $J_{\rm sp} = 2.2415(5)$],
$\delta=0.0025(2)$ in the $\phi^4$ model for $D=1/2$, 
$\delta= 0.0167(7)$ in the IsXY model with $C=0$.

\begin{figure}[tb]
\centerline{\psfig{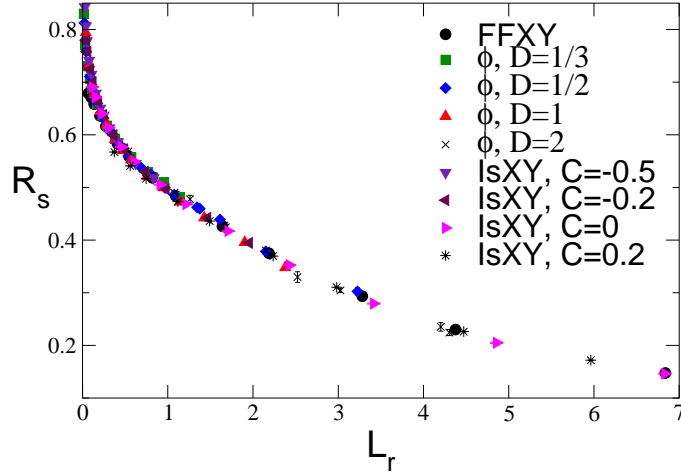}}
\caption{ 
$R_s$ at fixed $R_c$ versus the rescaled lattice size $L_r=L/l$.
We set $l=\xi_{s}^{(c)}\approx 118$ for the FFXY model.
}
\label{xiolr}
\end{figure}

\begin{figure}[tb]
\centerline{\epsfig{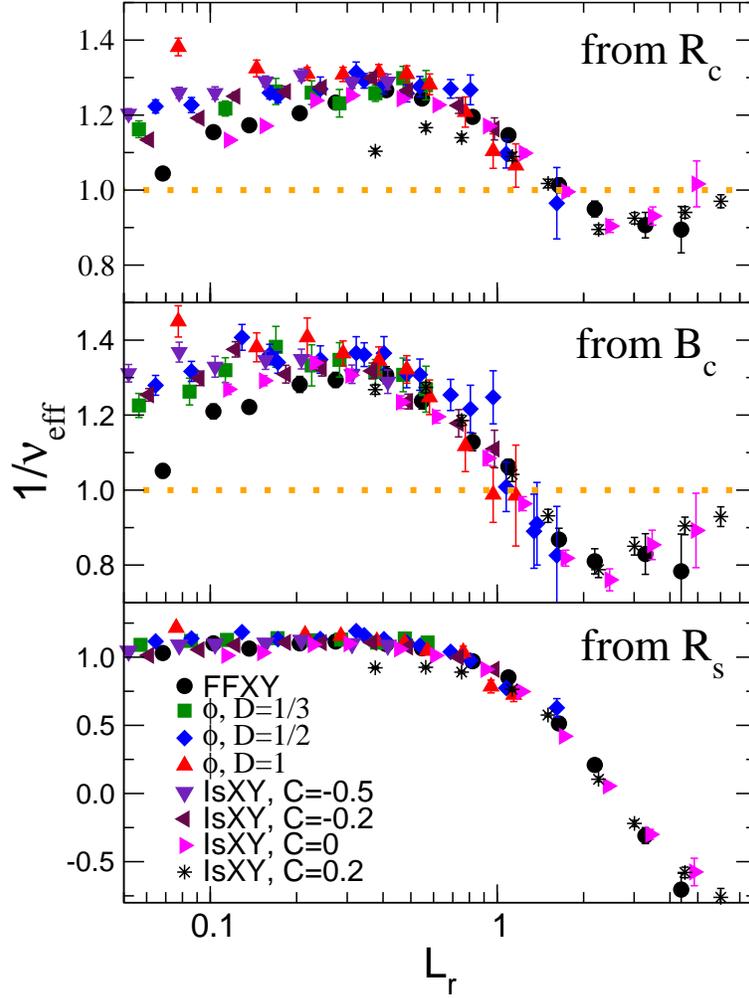}}
\caption{ 
$1/\nu_{\rm eff}$ computed by using $R_c$, $B_c$, and $R_s$ vs
$L_r=L/l$.  For $L_r\to\infty$, $1/\nu_{\rm eff}$ should converge to
$1/\nu_{\rm Is} = 1$ for $R_c$ and $B_c$, and $-1$ for $R_s$.  
}
\label{nueff}
\end{figure}

Our FSS analysis definitely shows that the chiral transition, when it
is continuous, belongs to the Ising universality class. However, the
Ising critical regime is reached only after a crossover region in
which effective exponents and RG invariant quantities show a behavior
that is surprisingly similar in the FFXY model, in the $\phi^4$ model
with $0\lesssim D\lesssim 2$, and in the IsXY model for $-0.5\lesssim
C \lesssim 0.2$.  Fig.~\ref{xiolr} shows the FSS curves of $R_s\equiv
\xi_s/L$ at fixed $R_c\equiv \xi_c/L$ plotted versus a rescaled
lattice size $L_{r} \equiv L/l$, where $l$ is a rescaling factor that
is chosen judiciously for each different model.  The optimal data collapse is
obtained for $l/l_{\rm FFXY} =4.6,3.2,1.8,1.1$ for the
$\phi^4$ model at $D=1/3,1/2,1,2$ and $l/l_{\rm FFXY}=2.6, 1.1,0.45,0.09$ for the
IsXY model at $C=-0.5,-0.2,0,0.2$. Of course, for $L$ large $\xi_s\to
\xi_s^{(c)}$ and therefore, given two different models, $l_1/l_2$
should correspond to the ratio of the corresponding spin correlation
lengths at the chiral transition. Thus, the scaling we observe implies
that $R_s$ is an approximately universal function of $L/\xi_s^{(c)}$.
Fig.~\ref{nueff} shows the effective
exponent $1/\nu_{\rm eff}$ defined by Eq.~(\ref{nueffdef}), as
obtained from $R_c$, $B_c$, and $R_s$, respectively, by using the same
rescaling factors $l$ as determined from $R_s$.  In all cases we
observe an approximate collapse of the data. Note that there is a
rather extended region, $L_r \lesssim 1$, i.e. $L \lesssim
\xi_s^{(c)}$, in which $\nu_{\rm eff}$ computed from the chiral
variables $R_c$ and $B_c$ (resp. $R_s$) is approximately 0.8
(resp. 0.9). This preasymptotic behavior explains previous estimates
$\nu\approx 0.8$ of the chiral exponent, see, e.g.,
\cite{GKLN-91,LGK-91,RJ-92,Lee-94,LL-94,KNKB-94,BG-97,LL-98,BD-98,LS-00,OI-03,GD-05}.
Actually, the data of $1/\nu_{\rm eff}$ from $R_c$ and $B_c$ appear to
undershoot the Ising value, which is then approached from below.  A
data collapse is also observed for $B_c$ itself,
the helicity modulus, and the effective exponent $\eta_{\rm eff}$
obtained from $\chi_c$.  In conclusion, the crossover to
the asymptotic Ising behavior presents interesting features and 
appears universal to some extent.

The origin of the apparent scaling behavior is not clear. 
It might reflect the nearby presence of a multicritical point, 
which could be observed only by performing a further fine
tuning of the Hamiltonian parameters, where chiral and spin modes
become critical at the same time. The identification of this 
multicritical point remains an open issue.

\begin{references}
\bibitem{afisxy}
In the IsXY model there is also 
an  antiferromagnetic Ising transition line in the region $C<-C_{\rm
Is}$, which starts from the point $C=-C_{\rm Is}$, $J=0$.
It is not shown in Fig.~\ref{phased}.
\bibitem{sccorr}
Scaling corrections are expected to be 
$O(L^{-\zeta})$, with $\zeta = 1/\eta-4$ for $\Upsilon$
and $\zeta= {\rm Min}(2-\eta,1/\eta-4)$ for $\chi_s$ and $R_s$.
\end{references}
\end{document}